\documentclass[letterpaper,journal]{IEEEtran}

\usepackage{amsmath,amsfonts,amssymb}
\usepackage{array}
\usepackage{booktabs}
\usepackage{cite}
\usepackage{graphicx}
\usepackage{url}

\begin{document}

\title{ReFLEX: Length-Generalizable CSI Denoising for MIMO-OFDM via Relative-Frequency Bias}

\author{Zhibin Zhang,
    Robert~N.~Potekhin,
    Ziwei Wan,~\IEEEmembership{Member,~IEEE},
    Vladimir Lyashev,~\IEEEmembership{Senior Member,~IEEE},
    and~Zhen Gao,~\IEEEmembership{Senior Member,~IEEE}

\thanks{Zhibin Zhang, Robert N. Potekhin, and Vladimir Lyashev are
with the Moscow Institute of Physics and Technology (State University),
Moscow, Russia (e-mails: zhibin@phystech.edu; potekhin.rn@phystech.edu;
lyashev.va@mipt.ru).}
\thanks{Ziwei Wan is with the Yangtze Delta Region Academy, Beijing Institute
of Technology (Jiaxing), Jiaxing 314019, China (e-mail:
ziweiwan@bit.edu.cn).}
\thanks{Zhen Gao is with the School of Interdisciplinary Science, Beijing
Institute of Technology, Beijing 100081, China (e-mail:
gaozhen16@bit.edu.cn).}
}


\maketitle

\begin{abstract}
This letter studies CSI denoising for MIMO--OFDM with variable NR resource block (RB) allocations. ReFLEX is a length-generalizable Transformer whose frequency attention uses a relative-frequency position bias (RFPB) generated from subcarrier offsets. A single checkpoint handles unseen RB lengths and can be applied to sparse DM-RS observations in the tested RB5/RB10 PUSCH setup without retraining. In a 3GPP~TR~38.901 UMa NLOS channel, ReFLEX achieves about $-9.6$~dB NMSE on unseen RB lengths. In NR PUSCH/UL-SCH simulations, ReFLEX denoising followed by time-frequency interpolation reduces the 10\% BLER threshold by about 2--3~dB.
\end{abstract}


\section{Introduction}

\IEEEPARstart{M}{IMO--OFDM} receivers require accurate channel state information (CSI) for equalization, detection, and decoding. Classical denoising and refinement methods include DFT-based truncation and LMMSE filtering. The performance of these methods depends on selecting a delay window as well as having knowledge of the channel covariance and noise variance\,\cite{van_de_beek1995ofdm,edfors1998svd}. Neural CSI estimators based on convolutional, residual, attention-based, and generative architectures have also been studied\,\cite{ye2018power,channelformer2023,cheng2023mstgan,zhou2026cdit}. Prior massive-MIMO CSI denoising work further showed that vector autoregression and tensor DnCNN models can improve noisy SRS-based channel estimates when combined with beam-delay-sparse representations\,\cite{artemasov2024autoregressive}.

In the reported setups, the closest related neural estimators remain tied to a fixed grid or preprocessing chain. Channelformer\,\cite{channelformer2023} applies self-attention to LS-based OFDM channel estimation and studies online adaptation, but its reported NR setup fixes $N_f=72$ subcarriers and the authors note that retraining is needed for significant changes such as the number of subcarriers. The beam-delay tensor DnCNN in\,\cite{artemasov2024autoregressive} targets SRS-based TDD massive-MIMO denoising and reports downlink SINR-oriented results. It first maps antenna-frequency CSI to a sparse beam-delay tensor using antenna-to-beam and frequency-to-delay DFT operations, applies a three-dimensional CNN denoiser, and then transforms the output back. This pipeline exploits sparsity, but it adds transform steps and was evaluated under fixed bandwidth, subcarrier spacing, and SRS-comb settings. Swin/cGAN-based estimation and diffusion inpainting techniques can exploit multi-slot pilot structures or handle noisy, sparse observations under varying pilot patterns and noise levels\,\cite{cheng2023mstgan,zhou2026cdit}. However, these forms of robustness alone do not demonstrate that a single checkpoint can process RB lengths that were absent during training.

This fixed-grid dependence matters in NR uplink scheduling, where RB allocations vary with traffic and bandwidth constraints\,\cite{3gpp38211}. Separate checkpoints for different RB lengths and SNR settings increase model-management cost, whereas fixed-length implementations generally cannot directly process untrained RB lengths. ReFLEX avoids relying on fixed frequency-position encoding tables by generating the attention bias from pairwise subcarrier offsets. This allows the same parameter set to be evaluated on the offset matrix induced by a new RB allocation.

We focus on RB-length generalization for CSI denoising and refinement from noisy channel estimates under fixed subcarrier spacing. The model operates on observed subcarrier positions (i.e., frequency-domain samples), which may be a dense observation set containing all allocated subcarriers or a sparse DM-RS observation set. When a receiver requires CSI on the data grid, the denoised sparse estimates are linearly interpolated across time and frequency.

We propose ReFLEX, a \emph{relative-frequency length-extensible} Transformer. ReFLEX preserves the frequency, polarization, and array dimensions of MIMO--OFDM CSI and applies axial attention along those dimensions. Its key mechanism is a relative-frequency position bias: the frequency-attention bias is generated as a continuous function of the relative subcarrier offsets rather than learned as a fixed absolute-position table. Under wide-sense stationary channel models, frequency correlation is more naturally tied to frequency separation than to an absolute subcarrier index. This structure allows the same RFPB parameters to generate a bias matrix for a new RB length or sparse observation set from the corresponding pairwise offsets.

Our contributions are: (i) we formulate and evaluate RB-length generalization for CSI denoising with a single parameter set; (ii) we propose ReFLEX combining tensor-preserving axial attention with continuous RFPB for variable RB lengths; and (iii) we report NMSE and BLER results on 3GPP UMa NLOS and NR PUSCH/UL-SCH scenarios (RB5/RB10).

\section{System Model and Problem Formulation}\label{sec2}

We consider per-user uplink CSI denoising and refinement on observed subcarrier positions. For an allocation of $R$ resource blocks (RBs) with $N_{\rm sc}=12R$ subcarriers, let $\mathcal S_R=\{s_1,\ldots,s_{M_R}\}\subseteq\{1,\ldots,N_{\rm sc}\}$ denote the observed subcarriers used as model input. Dense NMSE experiments use $M_R=N_{\rm sc}$, whereas a sparse DM-RS input has $M_R<N_{\rm sc}$. For compact notation, $\mathbf H_R\in\mathbb C^{M_R\times A\times P}$ denotes the normalized frequency-domain channel restricted to $\mathcal S_R$. Here $A=64$ is the number of elements in the $8\times8$ base-station antenna array and $P=2$ denotes dual polarization, yielding $AP=128$ receive ports. The channel is normalized by sample power so that errors remain comparable across RB lengths and channel realizations.

Let $\widetilde{\mathbf H}_R$ denote the normalized noisy input at the same observation locations:
\begin{equation}
\widetilde{\mathbf H}_R=\mathbf H_R+\mathbf E_R,
\end{equation}
where $\mathbf E_R \sim \mathcal{CN}(\mathbf 0,\sigma^2\mathbf I)$ models independent LS noise at the observation points. The denoiser estimates $\mathbf H_R$ on $\mathcal S_R$. When link-level detection requires CSI on PUSCH data resource elements, the denoised sparse estimates are linearly interpolated over time and frequency. The model input stacks the real and imaginary parts of $\widetilde{\mathbf H}_R$ as \begin{equation}
\mathbf X_R \in \mathbb R^{P\times M_R\times 2\times 8\times 8},
\end{equation} preserving the frequency, polarization, and two-dimensional array structure.

The model is trained on the RB-length set
$\mathcal R_{\rm tr}=\{2,4,6\}$, corresponding to 24, 48, and
72 subcarriers respectively. It is tested on
$\mathcal R_{\rm te}=\{2,3,4,5,6,8,10\}$, where the unseen
RB-length subset is
$\mathcal R_{\rm un}=\mathcal R_{\rm te}\setminus\mathcal R_{\rm tr}
=\{3,5,8,10\}$. A single parameter set is used for all test RB
lengths and for the sparse DM-RS BLER evaluation.

Let $\gamma$ denote the sample-level SNR. The estimator is
\begin{equation}
\widehat{\mathbf H}_R=f_{\boldsymbol\theta}(\widetilde{\mathbf H}_R,\gamma),
\end{equation}
where $\boldsymbol\theta$ is shared across RB lengths.  The RB length $R$ appears only through the observed subcarrier positions $\mathcal S_R$ and does not require a separate branch.  During training, the model uses dense subcarrier observations with
$\mathcal S_R=\{1,\ldots,N_{\rm sc}\}$. For RB5/RB10 PUSCH, the
dense-trained checkpoint is directly evaluated on sparse DM-RS
observations without additional retraining. The loss is the
sample-power-normalized mean-square error:
\begin{equation}
\mathcal L(\boldsymbol\theta)
=
\mathbb E_{R,\gamma}
\left[
\frac{
\left\|f_{\boldsymbol\theta}(\widetilde{\mathbf H}_R,\gamma)-\mathbf H_R\right\|_F^2
}{
\left\|\mathbf H_R\right\|_F^2
}
\right].
\end{equation}

CSI denoising performance is measured by
\begin{equation}
{\rm NMSE}
=
\mathbb E
\left[
\frac{
\left\|\widehat{\mathbf H}_R-\mathbf H_R\right\|_F^2
}{
\left\|\mathbf H_R\right\|_F^2
}
\right],
\end{equation}
reported as $10\log_{10}({\rm NMSE})$. Link-level impact is measured by block error rate (BLER) in a common PUSCH receiver chain where only the CSI estimate supplied to the equalizer is changed.

\section{Proposed ReFLEX}

\begin{figure*}[!t]
\centering
\includegraphics[width=\textwidth]{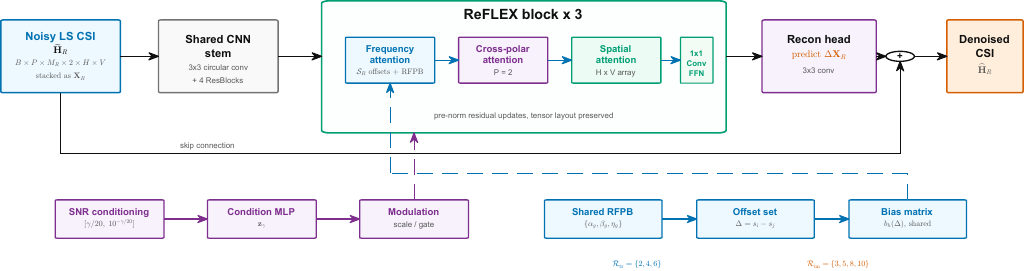}
\caption{Flowchart of ReFLEX. The CSI tensor structure is preserved while frequency, cross-polarization, and spatial axial attention are applied in each block. The shared RFPB is evaluated from the relative subcarrier offsets of the current observation set.}
\label{fig:architecture}
\end{figure*}

ReFLEX is designed for single-checkpoint CSI denoising with variable frequency-domain lengths. As shown in Fig.~\ref{fig:architecture}, it combines a tensor-preserving input and reconstruction path, axial attention over frequency, polarization, and spatial dimensions, and RFPB for RB-length generalization. All trainable parameters are shared across RB lengths; only the relative-offset matrix changes for a new RB length or sparse observation set.

\subsection{Overall Architecture}

For an input with $M_R$ observed subcarrier positions, the noisy CSI is arranged as
\begin{equation}
\mathbf X_R \in \mathbb R^{B\times P\times M_R\times 2\times H\times V},
\end{equation}
where $B$ is the batch size, $P=2$, $H\times V$ is the two-dimensional base-station antenna-array grid, and dimension $2$ contains the real and imaginary parts. A shared CNN stem extracts local array-plane features at each $(b,p,n)$ position and maps the input to
\begin{equation}
\mathbf F_0=f_{\rm stem}(\mathbf X_R),
\qquad
\mathbf F_0 \in \mathbb R^{B\times P\times M_R\times C_e\times H\times V}.
\end{equation}
Because the stem operates on the array plane, its parameters are independent of $M_R$.

ReFLEX uses the sample-level SNR as global side information.  The condition encoder maps $\gamma$ to a vector $\mathbf z_\gamma=\psi_\gamma([\gamma/20,\;10^{-\gamma/20}]^T)$ via a small MLP, encoding the normalized dB SNR and the relative noise level; this vector is injected into the input features and Transformer blocks.

The backbone stacks $D$ ReFLEX blocks, each applying frequency, polarization and spatial attention and a pointwise FFN.  The reconstruction head predicts a correction $\Delta \mathbf X_R$, and the output is $\widehat{\mathbf X}_R=\mathbf X_R+g(\mathbf z_\gamma)\Delta \mathbf X_R$, where $g$ is an SNR-dependent gate; real and imaginary components are recombined into $\widehat{\mathbf H}_R$.

\subsection{Tensor-wise Axial Transformer Block}

Instead of flattening CSI into a single token sequence, ReFLEX preserves the tensor
\begin{equation}
\mathbf F
\in
\mathbb R^{B\times P\times M_R\times C_e\times H\times V}
\end{equation}
throughout the backbone. For each fixed $(b,p,h,v)$, frequency attention processes the observed-frequency sequence $\{\mathbf F_{b,p,n,:,h,v}\}_{n=1}^{M_R}$ and uses RFPB to represent relative frequency distance. For each fixed $(b,n,h,v)$, cross-polarization attention exchanges information between the two polarization components. For each fixed $(b,p,n)$, spatial attention operates on the $HV$ array positions with a two-dimensional relative position bias.

With pre-normalization and residual connections, a block is written as
\begin{align}
\mathbf F' &=
\mathbf F
+
\mathcal A_f(\operatorname{Norm}(\mathbf F), \mathcal S_R),
\\
\mathbf F'' &=
\mathbf F'
+
\mathcal A_p(\operatorname{Norm}(\mathbf F')),
\\
\mathbf F''' &=
\mathbf F''
+
\mathcal A_s(\operatorname{Norm}(\mathbf F'')),
\\
\mathbf F_{\rm out} &=
\mathbf F'''
+
\operatorname{FFN}(\operatorname{Norm}(\mathbf F''')),
\end{align}
where $\mathcal A_f$, $\mathcal A_p$, and $\mathcal A_s$ denote frequency, polarization, and spatial attention. Only $\mathcal A_f$ depends on $\mathcal S_R$, and this dependence only determines the relative offsets used to generate the bias matrix.

\subsection{Relative-Frequency Position Bias}

RFPB is applied in frequency self-attention. For the $h$th attention head, the attention logit from observed subcarrier $s_i$ to observed subcarrier $s_j$ is
\begin{equation}
a_{ij}^{(h)}
=
\frac{
(\mathbf q_i^{(h)})^T \mathbf k_j^{(h)}
}{
\sqrt{d_h}
}
+
b_h(s_i-s_j),
\end{equation}
where $d_h$ is the per-head dimension, and the bias depends only on the relative subcarrier offset. Unlike absolute position encodings or fixed-length relative position tables~\cite{shaw2018relative}, RFPB generates the bias from a continuous function of $\Delta=s_i-s_j$.

ReFLEX organizes frequency-attention heads into triplets. Because the subcarrier spacing is fixed at $30$ kHz, the physical separation $\Delta f_{ij}=(s_i-s_j)\Delta f$ can be represented by $\Delta$ with a learnable scale. For the $g$th triplet, define
\begin{equation}
u_g(\Delta)=\alpha_g\Delta ,
\end{equation}
where $\alpha_g$ is a learnable frequency-scale parameter. The three bias curves are
\begin{align}
b_{3g}(\Delta)
&=
\eta_g
\left(
\frac{1}{1+u_g^2(\Delta)}
-
\beta_g
\right),
\\
b_{3g+1}(\Delta)
&=
\eta_g
\frac{u_g(\Delta)}{1+u_g^2(\Delta)},
\\
b_{3g+2}(\Delta)
&=
-b_{3g+1}(\Delta),
\end{align}
where $\alpha_g$, $\beta_g$, and $\eta_g$ form a learnable triplet. In implementation, $\alpha_g$ uses a softplus parameterization to ensure nonnegativity, $\beta_g$ uses a sigmoid parameterization and is constrained to $(0,1)$, and $\eta_g$ is a learnable scale. The first curve is even in $\Delta$ and favors nearby subcarriers, whereas the other two are odd and provide direction-sensitive offset features.

RFPB is motivated by OFDM frequency-domain correlation. Under the
wide-sense stationary uncorrelated scattering assumption and an
exponential power-delay profile, the frequency-domain correlation
is governed primarily by the subcarrier-index offset and can be written as
\begin{equation} \label{Practical_LMMSE}
r_f[\kappa]
=
\frac{1}{
1+j2\pi\tau_{\rm rms}\kappa\Delta f
},
\end{equation}
where $\kappa$ is the subcarrier-index offset, $\Delta f$ is the
subcarrier spacing, and $\tau_{\rm rms}$ is the RMS delay spread. Its real part is even and decays with $|\kappa|$, and its imaginary part is odd.  The RFPB triplet follows these symmetries as a structural prior rather than an estimate of the channel covariance.

Given a new observation set $\mathcal S'_R$, ReFLEX computes offsets $\Delta=s_i-s_j$ for $s_i,s_j\in\mathcal S'_R$ and uses the same $\{\alpha_g,\beta_g,\eta_g\}$ to generate the bias matrix, introducing no new RB-specific parameters.

\section{Numerical Results}

\subsection{Experimental Setup}\label{sec:setup}

Channels are generated by QuaDRiGa~\cite{QuaDRiGa} using the 3GPP TR 38.901 UMa NLOS scenario~\cite{3gpp38901}. The carrier frequency is $2.6$ GHz, the subcarrier spacing is $30$ kHz, and the base station uses an $8\times8$ dual-polarized antenna array. The RB train/test split follows Sec.~\ref{sec2}. Training SNRs are sampled uniformly from $[-20,20]$ dB, and test SNRs are $-20:4:20$ dB. ReFLEX uses $D=3$ axial Transformer blocks, embedding dimension $96$, and $6$, $2$, and $6$ heads for frequency, polarization, and spatial attention, respectively. The input stem contains one $3\times3$ circular convolution and $4$ residual blocks with SNR conditioning. The batch size is $8$, and the model is trained for $300$ epochs.

For the NMSE experiments, the baselines are a CNN-style estimator based on the NVIDIA Aerial channel-estimation~\cite{nvidia_aerial_ce}, frequency-only Practical LMMSE, and frequency-only Oracle-covariance LMMSE (Oracle-cov. LMMSE). Both LMMSE baselines are applied independently along subcarriers for each antenna/polarization port:
\begin{equation}
\mathbf h_{a,p,R}\in\mathbb C^{N_{\rm sc}},
\qquad
\widetilde{\mathbf h}_{a,p,R}
=
\mathbf h_{a,p,R}+\mathbf e_{a,p,R}.
\end{equation}
Given covariance $\mathbf R_f$ and noise variance $\sigma^2$, the estimate is
\begin{equation}
\widehat{\mathbf h}_{a,p,R}
=
\mathbf R_f
\left(
\mathbf R_f+\sigma^2\mathbf I
\right)^{-1}
\widetilde{\mathbf h}_{a,p,R}.
\end{equation}
Practical LMMSE constructs $\mathbf R_f$ using the exponential-PDP correlation in \eqref{Practical_LMMSE}, with $\tau_{\rm rms}$ set to the channel-generation RMS delay spread; it therefore assumes delay-spread side information but not sample covariance. Oracle-cov. LMMSE estimates $\mathbf R_f$ from clean CSI samples drawn from the test distribution and assumes known $\sigma^2$, making it a strong frequency-only linear reference rather than an absolute upper bound. These LMMSE references do not use spatial or polarization covariance.

The CNN baseline follows the default NVIDIA Aerial configuration with $32$ convolutional channels. To avoid penalizing a fixed-length CNN for RB or SNR mismatch, an independent CNN is trained for each test RB/SNR pair and denoted by CNN-RB/SNR. This gives an oracle-style fixed-length neural reference that knows the target RB size and operating SNR but requires multiple specialized checkpoints.

Link-level performance is evaluated using multiuser NR PUSCH/UL-SCH simulations~\cite{3gpp38211}. Channel estimation inputs are sparse LS estimates on the DM-RS time-frequency resource elements. The same ReFLEX checkpoint is applied per user at the observed DM-RS subcarrier positions, without retraining on sparse DM-RS patterns. The denoised sparse CSI is then linearly interpolated in time and frequency to the PUSCH resource grid used by the equalizer. The resulting CSI estimates are passed to the LMMSE equalizer for synchronous PUSCH data detection. The setup uses 12 single-layer UEs, 128 receive ports at the base station,
and 1000 slots per SNR point. The PUSCH configuration uses mapping type B,
DM-RS length $2$, additional position $1$, configuration type $2$, and
MCS index $10$ (16QAM, target code rate $0.6426$). To isolate CSI denoising,
all methods share the same MCS, LDPC decoder, noise generation, equalizer,
and receiver chain. The only changed component is the CSI estimate supplied
to the equalizer.

\subsection{NMSE on Unseen RB Lengths}

\begin{figure}[t]
    \centering
    \includegraphics[width=\linewidth]{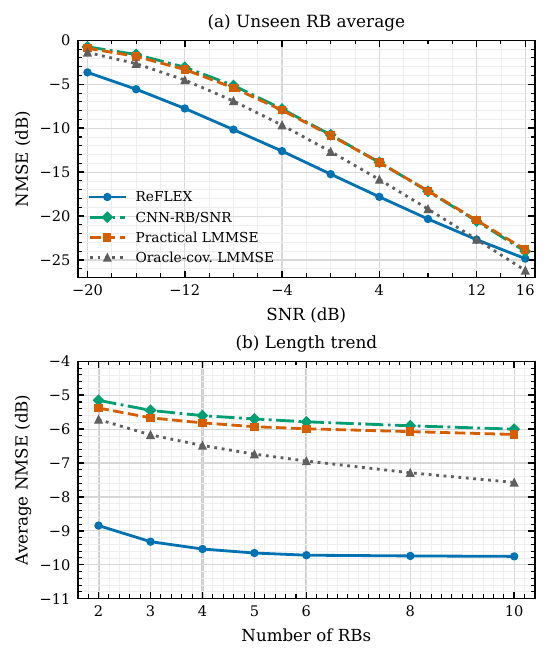}
    \caption{NMSE comparison under the 3GPP 38.901 UMa NLOS scenario. The RB train/test split follows Sec.~\ref{sec2}. (a) Average NMSE--SNR curves on unseen RB lengths. (b) Average NMSE across RB lengths; gray vertical lines indicate training RBs.}
    \label{fig:nmse}
\end{figure}

Fig.~\ref{fig:nmse}(a) reports NMSE versus SNR on unseen RB lengths. ReFLEX gives lower NMSE than the considered baselines at low and moderate SNRs. At $-20$ dB, ReFLEX reaches $-3.65$ dB, whereas CNN-RB/SNR, Practical LMMSE, and Oracle-cov. LMMSE obtain $-0.74$ dB, $-0.90$ dB, and $-1.38$ dB, respectively. At $0$ dB, ReFLEX reaches $-15.23$ dB, about $4.4$ dB lower than CNN-RB/SNR and Practical LMMSE. At high SNR, Oracle-cov. LMMSE benefits from test-distribution covariance and known noise variance and approaches or exceeds ReFLEX near the highest SNRs. At $16$ dB, ReFLEX obtains $-24.83$ dB and remains below CNN-RB/SNR and Practical LMMSE.

Fig.~\ref{fig:nmse}(b) shows average NMSE versus RB length. ReFLEX varies smoothly across seen RB2, RB4, and RB6 and unseen RB3, RB5, RB8, and RB10. The unseen-RB averages for RB3, RB5, RB8, and RB10 are $-9.32$ dB, $-9.65$ dB, $-9.74$ dB, and $-9.75$ dB, respectively. Although RB10 is longer than the maximum training length RB6, the NMSE does not degrade dramatically in this setup. CNN-RB/SNR and the LMMSE baselines yield higher average NMSE over the tested RB lengths.

\subsection{Ablation and Checkpoint Analysis}

Table~\ref{tab:ablation} summarizes the average NMSE over the tested
SNR values and RB lengths. ReFLEX achieves $-9.61$ dB on the unseen
RB-length subset, compared with $-5.76$ dB for CNN-RB/SNR, $-5.95$ dB
for Practical LMMSE, and $-6.91$ dB for Oracle-cov. LMMSE. Removing
RFPB degrades the unseen-RB average to $-7.47$ dB and the RB10 average
from $-9.75$ dB to $-7.22$ dB, which indicates that content-only
attention is less stable for cross-length frequency modeling.

The absolute-position-encoding (absolute PE) ablation uses a fixed
position table sized for RB10. Since the model is trained only on
$\mathcal R_{\rm tr}$, the position entries beyond RB6 are not directly
supervised. Absolute PE reaches $-9.46$ dB on seen RB lengths, close to
ReFLEX at $-9.35$ dB, but degrades to $-8.24$ dB on unseen RB lengths
and to $-6.84$ dB on RB10. This behavior is consistent with absolute
position indices tying the model to the training lengths. By contrast,
ReFLEX trained only on RB4 still reaches $-8.78$ dB on unseen RB lengths
and $-8.25$ dB on RB10, while multi-length training further improves
performance.

CNN-RB/SNR requires one checkpoint per RB/SNR operating point. Therefore,
a deployment covering $N_R$ RB lengths and $N_\gamma$ SNR levels requires
$N_RN_\gamma$ CNN checkpoints. ReFLEX uses a single 1.646M-parameter
checkpoint for all tested RB and SNR configurations.

\begin{table}[t]
\caption{Ablation and checkpoint comparison under the 3GPP 38.901 UMa NLOS scenario.}
\label{tab:ablation}
\centering
\small
\resizebox{\linewidth}{!}{%
\begin{tabular}{lccccc}
\toprule
Method & Train RB & Params & Seen avg. & Unseen avg. & RB10 avg. \\
\midrule
Prac. LMMSE & -- & -- & $-5.72$ & $-5.95$ & $-6.16$ \\
Oracle-cov. LMMSE & -- & -- & $-6.35$ & $-6.91$ & $-7.57$ \\
CNN-RB/SNR & RB/SNR & $N_RN_\gamma\times0.255$M & $-5.50$ & $-5.76$ & $-6.00$ \\
ReFLEX & 2,4,6 & 1.646M & $-9.35$ & $-9.61$ & $-9.75$ \\
w/o RFPB & 2,4,6 & 1.646M & $-7.62$ & $-7.47$ & $-7.22$ \\
Absolute PE & 2,4,6 & 1.657M & $-9.46$ & $-8.24$ & $-6.84$ \\
RB4 only & 4 & 1.646M & $-8.75$ & $-8.78$ & $-8.25$ \\
\bottomrule
\end{tabular}
}
\vspace{0.5ex}

\footnotesize{NMSE is reported in dB; lower is better. Seen and unseen averages are computed over the RB subsets defined in Sec.~\ref{sec2} and over test SNRs $-20:4:20$ dB.}
\end{table}

\subsection{Link-Level BLER}

\begin{figure}[t]
    \centering
    \includegraphics[width=\linewidth]{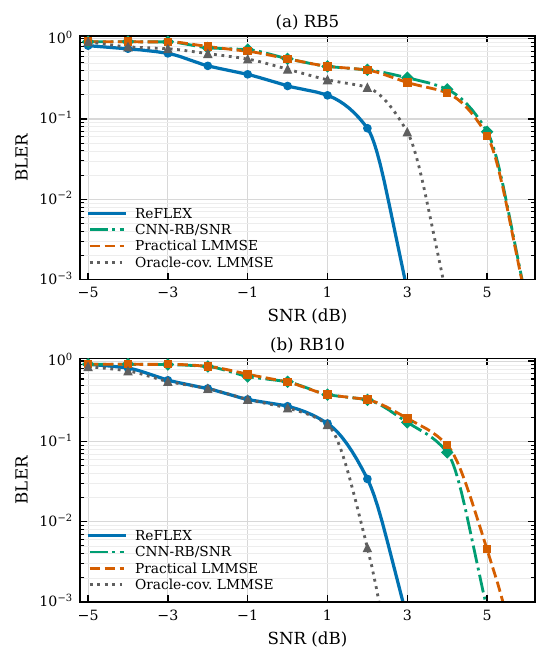}

    \caption{NR PUSCH BLER on unseen RB lengths when the estimated CSI is used by a multiuser LMMSE equalizer. Markers are simulated points, and curves are shape-preserving guides in the log-BLER domain. (a) RB5. (b) RB10. The dashed line denotes $10\%$ BLER; zero-error samples are omitted from the logarithmic axis.}
    \label{fig:bler}
\end{figure}

Fig.~\ref{fig:bler} reports coded-link performance on unseen RB5 and RB10 in the tested PUSCH setup.  For RB5, ReFLEX reaches BLER $7.67\times10^{-2}$ at $2$~dB, whereas CNN-RB/SNR and Practical LMMSE yield $4.10\times10^{-1}$ and $4.06\times10^{-1}$.  The corresponding 10\% BLER threshold is $1.72$~dB, giving about $2.9$~dB gains over CNN-RB/SNR and Practical LMMSE.

For RB10, ReFLEX achieves BLER $3.42\times10^{-2}$ at $2$~dB.  The corresponding 10\% BLER threshold is $1.32$~dB, within $0.2$~dB of Oracle-cov. LMMSE, and yields about $2.4$~dB gains over CNN-RB/SNR and Practical LMMSE.  Across RB5 and RB10, ReFLEX lowers the 10\% threshold by about 2--3~dB relative to CNN-RB/SNR and Practical LMMSE.

\section{Conclusion}

This letter studied RB-length generalization for MIMO-OFDM CSI denoising under fixed subcarrier spacing.  ReFLEX uses RFPB and axial attention to share one checkpoint across seen and unseen RB lengths.  In the 3GPP~TR~38.901 UMa NLOS setup, ReFLEX achieved about $-9.6$~dB average NMSE on unseen RB lengths and outperformed the Practical LMMSE and CNN-RB/SNR references.  In NR PUSCH/UL-SCH link-level simulations on RB5/RB10, ReFLEX denoised sparse DM-RS LS estimates and reduced the 10\% BLER threshold by 2--3~dB relative to these baselines.  Future work includes other DM-RS patterns, larger RB allocations, and multi-scenario training.

\end{document}